\documentclass[aip,jcp,twocolumn,reprint]{revtex4-1}

\usepackage{graphics,graphicx,amsfonts,amsmath,amsbsy,amssymb,color}
\usepackage{bm}
\usepackage[a4paper,vmargin={20mm,20mm},hmargin={20mm,10mm}]{geometry}
\usepackage{subfigure}
\usepackage{paralist}
\usepackage{bbold}
\usepackage{mathtools, nccmath}

\newcommand{\bra}{\langle}
\newcommand{\ket}{\rangle}
\newcommand{\bs}[1]{\boldsymbol{#1}}

\def \Nw {{N_{\textrm{w}}}}
\def \Nocc {{N_{\textrm{occ}}}}
\def \Nspawn {{N_{\textrm{spawn}}}}

\def \Ha {{\textrm{Ha}}}
\def \mHa {{\textrm{mHa}}}
\def \Eref {{E_{\textrm{ref}}}}
\def \Evar {{E_{\textrm{var}}}}
\def \EvarPT {{E_{\textrm{var+PT2}}}}

\DeclarePairedDelimiter{\nint}\lfloor\rceil
\DeclarePairedDelimiter{\abs}\lvert\rvert

\begin{document}

\title{A hybrid approach to extending selected configuration interaction and full configuration interaction quantum Monte Carlo}
\author{Nick~S.~Blunt}
\email{nicksblunt@gmail.com}
\affiliation{Department of Chemistry, Lensfield Road, Cambridge, CB2 1EW, United Kingdom}

\begin{abstract}
We present an approach to combining selected configuration interaction (SCI) and initiator full configuration interaction quantum Monte Carlo (i-FCIQMC). In the current i-FCIQMC scheme, the space of initiators is chosen dynamically by a population threshold. Here, we instead choose initiators as the selected space ($\mathcal{V}$) from a prior SCI calculation, allowing substantially larger initiator spaces for a given walker population. While SCI+PT2 adds a perturbative correction in the first-order interacting space (FOIS) beyond $\mathcal{V}$, the approach presented here allows a variational calculation in the same space, and a perturbative correction in the second-order interacting space. The use of a fixed initiator space reintroduces population plateaus into FCIQMC, but it is shown that the plateau height is typically only a small multiple of the size of $\mathcal{V}$. Thus, for a comparable fundamental memory cost to SCI+PT2, a substantially larger space can be sampled. The resulting method can be seen as a complementary approach to SCI+PT2, which is more accurate but slower for a common selected/initiator space. More generally, our results show that approaches exist to significantly improve initiator energies in i-FCIQMC, while still ameliorating the fermion sign problem relative to the original FCIQMC method.
\end{abstract}

\date{\today}

\maketitle

\section{Introduction}
\label{sec:intro}

Within electronic structure theory, several methods are available for systematic convergence to the exact solutions of a given system and basis set. One such option is full configuration interaction quantum Monte Carlo (FCIQMC)\cite{Booth2009, Spencer2012, Petruzielo2012, Booth2014}, a method which allows a stochastic sampling of the FCI solution by a projector Monte Carlo approach. In practical applications of FCIQMC, the initiator adaptation (i-FCIQMC) is almost always used\cite{Cleland2010, Cleland2011, Booth2011}, which biases the projection to ameliorate the sign problem, but can be converged to the exact solution with increasing walker populations. A second approach is selected configuration interaction (SCI or sCI)\cite{Huron1973, Buenker1974, Evangelisti1983, Scemama2013, Tubman2016, Holmes2016_2, Garniron2017, Schriber2016, Schriber2017}. In contrast to FCIQMC, SCI is a deterministic method which iteratively searches for the most important determinants in a desired solution, and performs a variational calculation in the selected space. It is then common to add a second-order perturbative correction to this SCI solution, referred to as SCI+PT2, which may be calculated by stochastic or semi-stochastic approaches\cite{Garniron2017, Sharma2017}. Similarly to i-FCIQMC, this method approaches exactness as the size of the selected space increases. These approaches are just two possibilities for systematic convergence to the FCI solution, with further approaches including the density matrix renormalization group (DMRG) method\cite{White1992, Chan2002, Chan2004, Olivares2015}, many-body expanded FCI (MBE-FCI)\cite{Eriksen2017, Eriksen2018, Eriksen2019}, and extensions to FCIQMC such as model space QMC (MSQMC)\cite{Ten-no2013, Ohtsuka2015, Ten-no2017}.

In the initial investigation of the FCIQMC approach, the entire FCI space was sampled without bias, giving an exact sampling of the desired solution. Performing FCIQMC in this way results in a population plateau in the simulation.\cite{Spencer2012} For walker populations less than this plateau, the ground state solution will die away compared to stochastic noise (the fermion sign problem), while above this plateau the solution may be sampled stably. As such, this plateau height sets a minimum memory requirement on the simulation. Crucially, the plateau height is often significantly lower than the size of the FCI space, allowing the exact solution to be sampled for systems where this state in unobtainable in deterministic approaches. Similarly, there will also be a population plateau if a fixed truncated space is sampled, at a height lower than the size of this space. Therefore, one approach to obtaining an accurate solution within this scheme would be to sample the largest truncated space possible for the available computational resources.

The original FCIQMC approach has since been improved upon by the initiator adaptation to FCIQMC (i-FCIQMC)\cite{Cleland2010, Cleland2011, Booth2011}. In this, the Hamiltonian is truncated by preventing spawnings between pairs of unimportant determinants. Here, the importance of a determinant is decided by a population criterion; all determinants with a population greater than $n_a$ are defined as initiators, and the Hamiltonian may act freely on these states. In contrast, non-initiators may only spawn to occupied determinants (i.e., all other Hamiltonian elements are set to $0$). Doing so removes the population plateau, so that an arbitrarily small walker population may be used. The number of initiators is always a small fraction of the walker population, so that the space sampled scales with the walker population and remains manageable. However, the fact that population plateaus are removed entirely from i-FCIQMC suggests that the truncation performed is actually too severe. For a given walker population there should exist an approximation for which more accurate results can be obtained, and where the sign problem remains manageable.

In this study we investigate one such approach, using a \emph{fixed} space of initiators taken as the selected space ($\mathcal{V}$) from a prior SCI calculation. It is important to recognize that this is not the same as performing the FCIQMC algorithm within $\mathcal{V}$, which would give back the same SCI result with lower efficiency. Rather, because initiators are allowed to spawn to any connected determinant, the space sampled (without rejections) in this approach contains both $\mathcal{V}$ and the space of its connections, known as the first-order interacting space (FOIS). The rejected spawnings then sample the \emph{second-order} interacting space, which may be used to estimate an accurate perturbative correction. Importantly, we will show that the plateau height for this approach is usually similar to the size of $\mathcal{V}$, even in cases where the FOIS is larger by several orders of magnitude.

This can be seen as an alternative approach to extending SCI, beyond the current SCI+PT2 protocol. Of course, another option to improving the accuracy in SCI+PT2 is to simply use a larger selected space. However, this option is only available when memory permits. SCI is eventually limited by the need to store a vector the size of $\mathcal{V}$ on each computing node. The approach presented here also allows variational estimates and reduced density matrices to be sampled without approximation\cite{Overy2014, Blunt2017} for the wave function in $\mathcal{V}$+FOIS, so that estimation of general observables is trivial. The approach also reduces the need for extrapolations to reach sub-milli-Hartree accuracy. In exchange for this improved accuracy, the approach to be described is certainly slower than SCI+PT2 methods for a common selected/initiator space. As such, in many cases the original SCI+PT2 approach will be preferable, and we instead see the method to be described as a complementary approach, which is particularly useful for treating very large active spaces with many orbitals. Variational Monte Carlo and diffusion Monte Carlo also exist as effective QMC approaches to extend SCI\cite{Scemama2013, Caffarel2016, Scemama2018, Scemama2018_2, Dash2018, PinedaFlores2019, Otis2019}, although these are quite different to the FCIQMC-based approach to be described.

It should be noted that selected spaces from SCI have been used as deterministic spaces in the semi-stochastic FCIQMC approach previously\cite{Petruzielo2012}. Since all deterministic states are automatically initiators in the semi-stochastic method, there is some similarity with the approach here. However, the approach to be presented goes beyond previous work in several ways. This includes using substantially larger selected spaces, similar in size to the walker population, the addition of a perturbative correction, and other advances to be described.

In Sec.~(\ref{sec:theory}) we discuss theory, providing a recap of FCIQMC and i-FCIQMC before describing the hybrid SCI/i-FCIQMC approach and the calculation of a PT2 correction. In Sec.~(\ref{sec:plateaus}) we investigate the sign problem in this hybrid approach by assessing population plateaus for various systems. The method is compared to the previous i-FCIQMC method in Sec.~(\ref{sec:ifciqmc_comp}), and to SCI+PT2 in Sec.~(\ref{sec:sci_comp}).

\section{Theory}
\label{sec:theory}

In this article we introduce a new approach to performing initiator FCIQMC calculations, which we denote i-FCIQMC(SCI). This approach performs i-FCIQMC, but with an initiator space that is \emph{fixed} to equal the selected space of a prior SCI calculation. This differs from the original i-FCIQMC approach, where the initiator space is non-constant and selected by a population criterion on the occupied walkers. Using SCI allows a substantially larger number of important determinants to be selected for the initiator space, thereby reducing the truncation placed on the Hamiltonian by the initiator approximation, for a given walker population. Spawned walkers rejected by the initiator criteria can additionally be used to construct an accurate PT2 correction to the remaining approximation. We begin by giving a recap of FCIQMC and its initiator approximation, before presenting this approach in detail.

\subsection{FCIQMC}
\label{sec:fciqmc}

FCIQMC\cite{Booth2009, Cleland2010, Cleland2011, Spencer2012, Petruzielo2012} is a projector QMC method, where the ground-state wave function is converged upon by repeated application of an operator $\hat{P} = \mathbb{1} - \Delta \tau (\hat{H} - E_S \mathbb{1})$ to an initial state,
\begin{equation}
| \Psi(\tau + \Delta \tau) \ket = | \Psi(\tau) \ket - \Delta \tau (\hat{H} - E_S \mathbb{1}) | \Psi(\tau) \ket,
\end{equation}
where $\hat{H}$ is the Hamiltonian operator, and $E_S$ is a shift applied to keep the walker population roughly constant. As has been pointed out\cite{Schwarz2017}, this can be derived as a steepest descent approach to minimize the variational energy, and performs imaginary-time propagation as in diffusion Monte Carlo and similar methods\cite{Umrigar1993, Foulkes2001}. Expanding the FCIQMC wave function in a basis of determinants (or other many-particle basis), $| \Psi(\tau) \ket = \sum_i C_i(\tau) | D_i \ket$, the coefficients are updated by
\begin{equation}
C_i(\tau + \Delta \tau) = C_i(\tau) - \Delta \tau \sum_j (H_{ij} - E_S \delta_{ij}) C_j(\tau).
\label{eq:itse}
\end{equation}
This projection could of course be performed exactly, which would obtain the FCI solution in the limit of large $\tau$. Instead, FCIQMC performs a stochastic sampling of both of the wave function coefficients, $C_i(\tau)$, and also the application of $\hat{H}$ in Eq.~(\ref{eq:itse}). Each iteration, each walker on $|D_j\ket$ randomly chooses a connected site $|D_i\ket$ (where $j \ne i$) to spawn to. The spawning amplitude is set such that the \emph{expected} amplitude spawned onto $|D_i\ket$ from $|D_j\ket$ is rigorously equal to $ - \Delta \tau H_{ij} C_j(\tau)$, ensuring that the algorithm is unbiased on average. For a determinant $|D_i\ket$ with amplitude $C_i(\tau)$, we define the number of walkers as $N_i$, each with amplitude $C_i(\tau)/N_i$. In the following we often drop the $\tau$ dependence of $\bs{C}(\tau)$ for notational ease. The full algorithm can then be achieved by the following steps:
\begin{enumerate}
\small{
\item \emph{Spawning:} Loop over all occupied determinants, $|D_j\ket$ (with $N_j$ walkers, each with a signed amplitude of $C_j/N_j$). For each walker on $|D_j\ket$, generate a connected determinant $|D_i\ket$ (where $i \ne j$ and $H_{ij} \ne 0$) with a probability $P_{\textrm{gen}}(i \leftarrow j)$, which should be calculated. Then create a new walker on $|D_i\ket$ with amplitude equal to $- \Delta \tau \frac{H_{ij}}{P_{\textrm{gen}}(i \leftarrow j)} \frac{C_j}{N_j}$.
\item \emph{Death:} Loop over all occupied determinants, $|D_i\ket$, and add $- \Delta \tau (H_{ii} - E_S) C_i$ to each amplitude.
\item \emph{Annihilation:} Merge all newly spawned walkers with the previous walkers to obtain the new walker list. Amplitudes on repeated determinants should be summed together, so that each occupied determinant appears only once in the list.
\item \emph{Rounding of small walkers:} For all determinants with $|C_i| < 1$, stochastically round the unsigned walker coefficient up to $1$ with probability $|C_i|$, otherwise kill the walker. Unoccupied determinants are not stored.
}
\end{enumerate}
To define the number of walkers, $N_i$, Umrigar and co-workers\cite{Petruzielo2012} suggest $N_i = \textrm{max}(1,\nint{\abs{C_i}})$, although other choices are possible\cite{fn1}.

The expected total amplitude spawned onto $| D_i \ket$ from all connected sites is given by
\begin{align}
S_i &= - \Delta \tau \sum_{j \ne i} H_{ij} C_j, \\
    &= - \Delta \tau \bra D_i | \hat{H}_{\textrm{off}} | \Psi \ket.
\label{eq:s_def}
\end{align}
Here, $\hat{H}_{\textrm{off}}$ contains the off-diagonal components of $\hat{H}$ in the FCIQMC basis. Note that diagonal elements of $\bs{H}$ are not included, as these are accounted for in the death step. Along with $\bs{C}$, $\bs{S}$ is the other large array stored in an FCIQMC implementation, and will be important for defining energy estimators later.

There are other propagators which can be applied to perform a stochastic sampling of the ground-state wave function. We recently suggested the use of preconditioning in FCIQMC\cite{Blunt2019}, which uses the following projection:
\begin{equation}
C_i(\tau + \Delta \tau) = C_i(\tau) - \frac{\Delta \tau}{H_{ii} - E} \sum_{j} ( H_{ij} - E \delta_{ij} ) C_j(\tau).
\label{eq:precond}
\end{equation}
This allows the time step to be set much larger, such that FCIQMC can typically converge within $20-30$ iterations. However, the expense of a step in FCIQMC also scales closely with the step size, such that the overall time savings for convergence are limited (although excitation generators optimized for the preconditioned case could change this). Nonetheless, we showed that this is a substantial benefit for the calculation of perturbative estimators, and so we will often use this propagation in our results. Note that the energy $E$ in Eq.~(\ref{eq:precond}) is set equal to the value of the projected energy, $\Eref$, from the current iteration:
\begin{align}
\Eref &= \frac{ \bra D_0 | \hat{H} | \Psi \ket }{ \bra D_0 | \Psi \ket }, \\
      &= \frac{ \sum_j H_{0j} C_j }{ C_0 },
\label{eq:hf_estimator}
\end{align}
which ensures that the population on the $|D_0\ket$, the reference determinant, remains exactly constant.

\subsection{i-FCIQMC}
\label{sec:i-fciqmc}

It may appear that the above algorithm can be performed with arbitrarily low walker populations. However, in practice a sign problem is encountered below a system-specific walker population, which effectively sets a minimum memory requirement for the simulation. As the walker population grows in the early stages of a simulation, the population eventually enters a `plateau' period; if one chooses to stabilize the population before this point then a severe sign problem is observed.

To overcome this, Cleland \emph{et al}. introduced the initiator adaptation to FCIQMC (i-FCIQMC)\cite{Cleland2010, Cleland2011}. In this, initiators are defined as determinants with a population ($|C_i|$) greater than a threshold, $n_a$, which is usually set to either $2$ or $3$. During propagation, certain Hamiltonian elements are then effectively set to $0$ by preventing spawning between certain determinants. Specifically:
\begin{enumerate}
\item Initiators may spawn to any connected determinant.
\item Non-initiators may only spawn to already-occupied determinants. Other spawned walkers generated from non-initiators are rejected.
\end{enumerate}
In the original scheme\cite{Cleland2010}, spawnings from two non-initiators to the same site with the same sign were also accepted, even if previously unoccupied. This is referred to as the `coherent-spawning rule', which we do not use in this article.

\subsection{i-FCIQMC(SCI)}
\label{sec:sci_iqmc}

In the above scheme, initiators are defined as determinants with walker populations greater than $n_a$. Because the vast majority of occupied determinants have exactly $|C_i|=1$, the size of the initiator space is always a small fraction of the total walker population. In practice this ensures that the plateau height is much smaller than the current walker population, so that population plateaus are not encountered. However, this can be viewed as evidence that the walker population in use is unnecessarily large for the given truncation applied to $\hat{H}$, or equivalently that a less severe truncation could be applied with the same walker population.

It would therefore be sensible to investigate approaches to increase the size of the initiator space for a given walker population. For this purpose, selected CI (SCI) is perhaps ideally suited. SCI is a deterministic method which iteratively selects the most important determinants in a solution according to some criterion.\cite{Huron1973, Buenker1974, Evangelisti1983, Scemama2013, Tubman2016, Holmes2016_2, Garniron2017} Each iteration, SCI diagonalizes the Hamiltonian within a selected space, $\mathcal{V}$, (sometimes referred to as the `reference' or `variational' space), and then expands this space by choosing the most important connected determinants by some criteria, repeating this process until convergence. Several variants of SCI exist, often with differing selection criteria. In the CIPSI (configuration interaction using a perturbative selection done iteratively) method\cite{Huron1973, Evangelisti1983} the selection criteria is obtained by first-order perturbation theory,
\begin{equation}
f^{\textrm{CIPSI}}_i = \Big| \frac{\sum_j H_{ij} C_j}{E_0 - H_{ii}} \Big|,
\end{equation}
where $E_0$ is the energy within the current selected space. Meanwhile, HCI (heat-bath configuration interaction)\cite{Holmes2016_2, Sharma2017, Li2018} uses the following metric,
\begin{equation}
f^{\textrm{HCI}}_i = \textrm{max}_j (|H_{ij} C_j|).
\end{equation}
Each iteration, the selected space is expanded to include all connected determinants with $f_i$ greater than some threshold, $\epsilon$. As $\epsilon \to 0$, the algorithm approaches FCI. This SCI procedure provides an effective way to select the most important determinants (those with the largest amplitudes in the FCI wave function), up to a truncation of some desired size.

To improve upon the approximation in i-FCIQMC, we therefore suggest defining the initiator space to equal the selected space from a prior SCI calculation. The initiator space is then \emph{fixed}, with no initiators arising from the usual population criterion. Besides this, the initiator rules are almost unchanged; walkers within the initiator space may spawn to any connected determinant, occupied or otherwise, while non-initiators may spawn to any occupied determinant. However, we add the rule that a non-initiator may spawn to any determinant within the initiator space, \emph{even if it is not occupied}. This latter situation never occurs in the original approach, but may occur in this modified method.

For ease of referring to this approach rather than i-FCIQMC or SCI+PT2 methods, we will use the acronym i-FCIQMC(SCI). Here, the acronym in parentheses denotes the method used to construct the initiator space (one could imagine using other methods for this purpose). As usual, `i-FCIQMC' will refer to the original initiator FCIQMC method, defining initiators by the population criterion. 

We note one exception to the above (which has been used for the results in this article, but which is non-essential): if an initiator spawns to an unoccupied determinant, then all other spawning attempts to that determinant (in the same iteration) are accepted, even from non-initiators. In this article, this exception is made both for i-FCIQMC and i-FCIQMC(SCI) results.

It is important to consider what this approach does from a theoretical point of view. Evolving by Eqs.~(\ref{eq:itse}) or (\ref{eq:precond}) will converge to the ground state of the corresponding Hamiltonian. Thus, understanding the accuracy of this approach requires understanding the truncation on $\hat{H}$.

It can be seen that this hybrid approach interpolates between two Hamiltonian truncations as the walker population is increased. We denote these two Hamiltonians $\bs{H}_A$ and $\bs{H}_B$:
\begin{equation}
\bs{H}_A = \begin{pmatrix}
           \bs{H}_{\mathcal{V}}  & \bs{H}_{\mathcal{VC}}^{\dagger} \\
           \bs{H}_{\mathcal{VC}} & \bs{D}_{\mathcal{C}} \\
            \end{pmatrix}
\end{equation}
\begin{equation}
\bs{H}_B = \begin{pmatrix}
           \bs{H}_{\mathcal{V}}  & \bs{H}_{\mathcal{VC}}^{\dagger} \\
           \bs{H}_{\mathcal{VC}} & \bs{H}_{\mathcal{C}} \\
            \end{pmatrix}.
\end{equation}
Here, $\mathcal{V}$ is the selected space and $\mathcal{C}$ is the space of all connections to $\mathcal{V}$, excluding those already in $\mathcal{V}$ (the FOIS). $\bs{H}_{\mathcal{V}}$ is the Hamiltonian projected into $\mathcal{V}$, $\bs{H}_{\mathcal{C}}$ is the Hamiltonian projected into $\mathcal{C}$, and $\bs{D}_{\mathcal{C}}$ is the Hamiltonian diagonal in $\mathcal{C}$. $\bs{H}_{\mathcal{VC}}$ contains elements connecting $\mathcal{V}$ and $\mathcal{C}$. We have ignored Hamiltonian elements beyond $\mathcal{C}$ here, as they do not contribute to this discussion.

It can be seen that i-FCIQMC(SCI) interpolates between these two Hamiltonians as follows. Walkers in $\mathcal{V}$ (which are initiators) may always spawn to any connected determinant in either $\mathcal{V}$ or $\mathcal{C}$, corresponding to $\bs{H}_{\mathcal{V}}$ and $\bs{H}_{\mathcal{VC}}$ blocks, respectively. However, in the limit of a zero walker population, walkers in $\mathcal{C}$ will be unable to spawn to any determinants except those in $\mathcal{V}$, since those in $\mathcal{C}$ will be both unoccupied and non-initiators. This corresponds to spawning with $\bs{H}_A$. In contrast, in the limit of large walker populations, all determinants in $\mathcal{C}$ will be occupied, so that all walkers may spawn freely between and within $\mathcal{V}$ and $\mathcal{C}$. This corresponds to spawning with $\bs{H}_B$. Therefore, the hybrid i-FCIQMC(SCI) approach will interpolate between these two Hamiltonians as the walker population is increased. An off-diagonal element of $\bs{H}_{\mathcal{C}}$ will be set to zero whenever a corresponding determinant is unoccupied. We will assess the accuracy of this approach with varying walker population in Sec.~(\ref{sec:plateaus}), and compare to SCI+PT2 in Sec.~(\ref{sec:sci_comp}).

It is reasonable to ask whether it is better to simply use either $\bs{H}_A$ or $\bs{H}_B$ - what is the benefit of the intermediate Hamiltonian that depends on the instantaneous walker distribution? Ideally we would like to use $\bs{H}_B$, which is the full Hamiltonian projected into $\mathcal{V} \oplus \mathcal{C}$. However, doing so requires knowing whether a newly-spawned walker is within this space. This either requires storing $\mathcal{C}$, which is not feasible, or checking whether the new determinant is connected to any state in $\mathcal{V}$, which is far too expensive, even with approaches for quickly finding connections as described in Ref.~(\onlinecite{Li2018}). Moreover, we have tested this Hamiltonian and find that the resulting population plateau is substantially higher than with the proposed Hamiltonian choice, often by several orders of magnitude, even though variational energies are not substantially lower. In contrast, $\bs{H}_A$ is simple and efficient to use, but is less accurate than the proposed i-FCIQMC(SCI) Hamiltonian, and does not reduce the population plateau noticeably. Moreover, we find that the PT2 correction (to be described) is less effective, even after updating the form of this correction appropriately. Thus, the above initiator rules are a pragmatic choice, which give accurate results within $\mathcal{V} \oplus \mathcal{C}$ for reasonable memory requirements, as will be demonstrated.

Note that while most determinants in $\mathcal{C}$ will be instantaneously unoccupied, every determinant in both $\mathcal{V}$ and $\mathcal{C}$ has the opportunity to become occupied during the simulation, and final estimates will be an average over all such states. Despite this, the average wave function will \emph{not} equal the true ground state of $\bs{H}_B$, and variational energies will be higher than the corresponding energy.

We also note that energies obtained from the solution of $\bs{H}_A$ are closely related to those from second-order perturbation theory, as performed in SCI+PT2. Indeed, this eigenvalue problem for $\bs{H}_A$ can be recast as an effective Hamiltonian eigenvalue problem entirely in $\mathcal{V}$, as has been performed by Loos and co-workers\cite{Garniron2018}. However, the final method here will differ, primarily by allowing trivial construction of a variational energy estimate (and reduced density matrices) for the resulting wave function, and perturbative corrections in the second-order space.

Lastly we point out that while the final result has some dependence on simulations parameters, this dependence is less than in the current i-FCIQMC scheme where the initiator space itself is dependent on parameters such as the time step and excitation generator. This is most notable in the case of bloom events, where a spawned walker is created with a magnitude greater than, $n_a$, instantly becoming an initiator. Using a fixed initiator space removes this dependence.

\subsection{Estimators and i-FCIQMC(SCI)+PT2}
\label{sec:sci_iqmc_pt2}

We will now describe how to calculate a second-order perturbative (PT2) correction from the rejected spawnings in the above approach. This is similar to the PT2 correction in SCI+PT2, but will sample a much larger space for a given $\mathcal{V}$ - up to the second-order interacting space beyond $\mathcal{V}$. It is also the same estimator that has been applied recently to the previous i-FCIQMC method\cite{Blunt2018, Blunt2019}, but will now sample a larger space for a given walker population.

We first describe how to calculate a variational energy estimate, $\Evar$, for the sampled wave function. Variational energies were originally obtained in FCIQMC through the two-body reduced density matrix (2-RDM)\cite{Overy2014, Blunt2017}. However, the accumulation of 2-RDMs can slow a simulation down considerably. Instead, we show how to calculate $\Evar$ without the 2-RDM, as first suggested in Ref.~(\onlinecite{Blunt2019}). We do not calculate RDMs for any results in this article, although it is straightforward if desired.

As described in Section~\ref{sec:i-fciqmc}, the two main arrays stored in an FCIQMC implementation are the walker array itself, with coefficients $\bs{C}$, and the spawned walker array, with coefficients $\bs{S}$, as defined in Eq.~(\ref{eq:s_def}). After communication, $\bs{C}$ and $\bs{S}$ are distributed across processes based on their indices by the same mapping. Thus, $C_i$ and $S_i$ are always stored on the same process for any $|D_i\ket$, so that estimators expressed as products of $C_i$ and $S_i$ with same index are efficient to calculate.

The variational energy, $\Evar$, may then be obtained as
\begin{align}
\Evar &= \frac{ \bra \Psi^1 | \hat{H} | \Psi^2 \ket }{ \bra \Psi^1 | \Psi^2 \ket }, \\
      &= \frac{ \sum_{ij} C_i^1 H_{ij} C_j^2 }{ \sum_i C_i^1 C_i^2 }, \\
      &= \frac{ \sum_i C_i^1 [ -S_i^2 / \Delta \tau + H_{ii} C_i^2 ] }{ \sum_i C_i^1 C_i^2 }.
\end{align}
Here, replica sampling is used to avoid biases, as described elsewhere\cite{Zhang1993, Overy2014, Blunt2014, Blunt2015_2}. Specifically, $|\Psi^1\ket = \sum_i C_i^1 |D_i\ket$ and $|\Psi^2\ket = \sum_i C_i^2 |D_i\ket$ are obtained from two independent FCIQMC simulations with different initial random number seeds. This avoids biases in estimating expectation values of products of random variables. In practice $\Evar$ and its statistical error are estimated by averaging the numerator and denominator of the above expression over all iterations after convergence (which is also the case for all estimators). Spawnings from both replicas can be used to reduce the noise in this expression, giving the final estimator used,
\begin{equation}
\Evar = \frac{ \sum_i C_i^1 H_{ii} C_i^2 - \frac{1}{2\Delta\tau} \sum_i [ C_i^1 S_i^2 + S_i^1 C_i^2 ] }{ \sum_i C_i^1 C_i^2 }.
\label{eq:var_energy}
\end{equation}

We next discuss how to calculate a PT2 correction to $\Evar$ from the rejected spawned walkers. This estimator was originally derived by supposing that i-FCIQMC samples a zeroth-order space, so that a PT2 correction can be constructed by analogy with SCI+PT2.\cite{Blunt2018} However, a more careful derivation can be given by using the approach taken by Guo \emph{et al.}\cite{Guo2018_2} in the context of DMRG\cite{White1992, Chan2002, Chan2004, Olivares2015} (and a similar derivation by Sharma\cite{Sharma2018}). Specifically, we define a zeroth-order Hamiltonian as
\begin{equation}
\hat{H}_0 = \hat{P} E_0 \hat{P} + \hat{Q} \hat{H}_{\textrm{d}} \hat{Q},
\end{equation}
where $\hat{P} = | \Psi \ket \bra \Psi |$, $\hat{Q} = \mathbb{1} - \hat{P}$,  $|\Psi\ket$ is the FCIQMC wave function and $E_0$ is a corresponding energy estimate (which will be set to $\Evar$). $\hat{H}_{\textrm{d}}$ contains the diagonal elements of $\hat{H}$ in the FCIQMC basis. With this definition of $\hat{H}_0$, the FCIQMC wave function and energy are, by construction, the zeroth-order wave function and energy, and the first-order correction to the energy is zero. The second-order correction is obtained by
\begin{equation}
\Delta E_2 = - \bra \Psi | \hat{V} \hat{Q} [ \hat{H}_0 - E_0 ]^{-1} \hat{Q} \hat{V} | \Psi \ket,
\end{equation}
where $\hat{V} = \hat{H} - \hat{H}_0$. This expression can be evaluated exactly, as was done by Guo \emph{et al.}\cite{Guo2018_2} We simply state their result, using the same notation,
\begin{equation}
\Delta E_2 = A + \frac{|B|^2}{C},
\end{equation}
where
\begin{equation}
A = - \sum_i \frac{ | \bra \Psi | \hat{V} \hat{Q} | D_i \ket |^2 }{ H_{ii} - E_0 },
\end{equation}
\begin{equation}
B = \sum_i \frac{ \bra \Psi | \hat{V} \hat{Q} | D_i \ket \bra D_i | \Psi \ket }{ H_{ii} - E_0 },
\end{equation}
\begin{equation}
C = \sum_i \frac{ | \bra D_i | \Psi \ket |^2 }{ H_{ii} - E_0 }.
\end{equation}
It was shown that the term $|B|^2/C$ is essentially negligible compared to $A$. We have implemented this term in the i-FCIQMC(SCI) approach and find the same conclusion, and so we do not consider this further. We instead take $\Delta E_2 = A$. To write this term in a form accessible to the current method, first note that for the definition of $\hat{H}_0$ given above, $\hat{Q} \hat{V} | \Psi \ket = (\hat{H} - E_0) | \Psi \ket$. Therefore,
\begin{equation}
\Delta E_2 = - \sum_i \frac{ | \bra \Psi | (\hat{H}_{\textrm{off}} + \hat{H}_{\textrm{d}} - E_0) | D_i \ket |^2 }{ H_{ii} - E_0 }.
\end{equation}
The replica trick can be used to avoid significant bias, setting $\bra \Psi | \rightarrow \bra \Psi^1 |$ and $ | \Psi \ket \rightarrow | \Psi^2 \ket$. Then,
\begin{align}
\Delta E_2 &= - \sum_i \frac{1}{ H_{ii} - E_0 } \bra \Psi^1 | (\hat{H}_{\textrm{off}} + \hat{H}_{\textrm{d}} - E_0) | D_i \ket \nonumber \\
  &\;\;\;\;\;\;\;\;\;\;\;\;\;\;\;\;\;\;\;\;\;\;\;\;\; \times \bra D_i | (\hat{H}_{\textrm{off}} + \hat{H}_{\textrm{d}} - E_0) | \Psi^2 \ket, \label{eq:A_1} \\[10pt]
  &= - \sum_i \frac{1}{ H_{ii} - E_0 } \Big[ -\frac{S_i^1}{\Delta \tau} + (H_{ii} - E_0)C_i^1 \Big] \nonumber \\
  &\;\;\;\;\;\;\;\;\;\;\;\;\;\;\;\;\;\;\;\;\;\;\;\; \times\Big[ -\frac{S_i^2}{\Delta \tau} + (H_{ii} - E_0)C_i^2  \Big], \label{eq:A_2} \\[12pt]
  &= \frac{1}{ (\Delta \tau)^2 } \sum_i \frac{ S_i^1 S_i^2 }{ E_0 - H_{ii} } + \frac{1}{\Delta\tau} \sum_i (S_i^1 C_i^2 + C_i^1 S_i^2) \nonumber \\
  &\;\;\;\;\;\;\;\;\;\;\;\;\;\;\;\;\;\;\;\;\;\;\;\; + \sum_i (E_0 - H_{ii}) C_i^1 C_i^2, \label{eq:A_3} \\[5pt]
  &= \frac{1}{ (\Delta \tau)^2 } \sum_i \frac{ S_i^1 S_i^2 }{ E_0 - H_{ii} } - \sum_i (E_0 - H_{ii}) C_i^1 C_i^2. \label{eq:A_4}
\end{align}
To get from Eq.~(\ref{eq:A_3}) to (\ref{eq:A_4}), we took the zeroth order energy ($E_0$) to equal the variational energy ($\Evar$) and used Eq.~(\ref{eq:var_energy}). This gives an expression for the (unnormalized) PT2 correction to $\Evar$.

It is also possible to obtain a simplified form for $A$ which typically has smaller noise than Eq.~(\ref{eq:A_4}), and will be our final expression. To obtain this, we can divide determinants $|D_i\ket$ into two groups depending on whether spawnings are rejected (in both replicas) or allowed (in at least one replica) by the initiator criteria:
\begin{align}
A &= \frac{1}{ (\Delta \tau)^2 } \sum_{i \, \in \, \textrm{rejected}} \frac{ S_i^1 S_i^2 }{ E_0 - H_{ii} } \nonumber \\
  &+ \frac{1}{ (\Delta \tau)^2 } \sum_{i \, \in \, \textrm{allowed}} \frac{ S_i^1 S_i^2 }{ E_0 - H_{ii} } - \sum_{i \, \in \, \textrm{allowed}} (E_0 - H_{ii}) C_i^1 C_i^2. \label{eq:A_5}
\end{align}
When the simulation is in equilibrium we have that $C_i(\tau + \Delta \tau) = C_i(\tau)$, on average. Using Eq.~(\ref{eq:itse}) or (\ref{eq:precond}) and rearranging then gives
\begin{align}
C_i &= \frac{1}{E - H_{ii}} \sum_{j \ne i} H_{ij} C_j, \\
    &= - \frac{ 1 }{ \Delta \tau } \frac{ S_i }{ E - H_{ii} }, \label{eq:equilib}
\end{align}
once the simulation has equilibrated. This is only true for determinants $|D_i\ket$ where the spawnings are \emph{not} rejected due to the initiator criteria, otherwise Eq.~(\ref{eq:itse}) and (\ref{eq:precond}) do not apply. As a result, the second and third terms on the right hand side of Eq.~(\ref{eq:A_5}) cancel on average.
%This assumes that $E_0$ in Eq.~(\ref{eq:A_5}) and $E$ in Eq.~(\ref{eq:equilib}) are equal, which will not be the case in general, but will always be approximately true.

Therefore the final expression for the PT2 correction is
\begin{equation}
\Delta E_2 = \frac{1}{ (\Delta \tau)^2 } \sum_{i \, \in \, \textrm{rejected}} \frac{ S_i^1 S_i^2 }{ E_0 - H_{ii} },
\label{eq:e_pt2}
\end{equation}
which applies equally for both the previous and current initiator FCIQMC approaches. This estimate of $\Delta E_2$ is formed as a sum over spawnings which are rejected due to the initiator criteria. There must be such a rejection on both replicas $1$ and $2$ on the same determinant for a non-zero contribution. This correction must be added to the variational energy estimate, $\Evar$, to be accurate. However, any reasonable estimate of the ground-state energy can be used in the denominator of Eq.~(\ref{eq:e_pt2}), since $E_0$ is well separated from all contributing $H_{ii}$.

Because $\Delta E_2$ is constructed from spawned walkers already present in a simulation, it can be constructed almost for free, requiring no significant extra computational steps. This is a notable difference to SCI+PT2, where the PT2 calculation is a separate algorithmic step.

\subsection{Summary of procedure}
\label{sec:procedure_summary}

As well as using SCI to generate the initiator space, we also use orbitals optimized by SCI. This approach was described by Smith \emph{et al.},\cite{Smith2017} and has become common in performing heat-bath CI calculations recently\cite{Chien2018, Li2018}. The orbitals are obtained by performing a CASSCF procedure to optimize active-active rotations, using SCI as an approximate solver within the active space. In cases where FCI is desired, the CAS is simply chosen as the entire space. In cases where one wishes to perform a CASSCF calculation within a reduced active space, active-active rotations can be optimized together with core-active, core-virtual and active-virtual rotations. We take this approach in all results presented, unless stated otherwise. When heat-bath CI (HCI) is used as the SCI method, Smith \emph{et al.} refer to this approach as aHCISCF. Below we refer to the more general approach as SCI-CASSCF.

The procedure taken in this article can be summarized as follows:
\begin{enumerate}
\item Optimize the molecular orbitals by performing a SCI-CASSCF procedure with a fixed SCI threshold.
\item Using these optimized orbitals, perform a further SCI calculation (with a different threshold, in general) and output the final selected space determinants.
\item Perform an initiator FCIQMC simulation, defining the space of initiators as the selected space from SCI. The initiator space is fixed throughout the simulation. Typically, we also initialize the FCIQMC wave function as the SCI wave function.
\item We consider estimators $\Evar$, constructed from Eq.~(\ref{eq:var_energy}), and $\Evar + \Delta E_2$, constructed from Eq.~(\ref{eq:e_pt2}).
\end{enumerate}
For all of our i-FCIQMC(SCI) calculations the SCI method is taken as HCI, but any SCI method would be valid. We also use the semi-stochastic adaptation to FCIQMC\cite{Petruzielo2012, Blunt2015}, where part of the FCIQMC projection is performed exactly within a deterministic space. For i-FCIQMC(SCI), we always take the deterministic space to equal the selected space, unless stated otherwise. Previously, we typically used deterministic spaces of size $\sim 10^4$, and never more than $10^6$. In the following we will sometimes use much larger deterministic spaces for the semi-stochastic approach, up to $\sim 2 \times 10^7$ in the current work. To aid in this task, we have implemented and used the algorithm for fast Hamiltonian construction described recently by Li \emph{et al.}\cite{Li2018}

\section{Results}
\label{sec:results}

All of our SCI and SCI+PT2 calculations were performed using the semi-stochastic heat-bath CI (SHCI) method, using the Dice code\cite{Dice}. The PySCF package\cite{pyscf} was used to generate molecular integral files, and also to perform CASSCF optimization of orbitals, using the algorithm described in Ref.~(\onlinecite{sun2017}). FCIQMC calculations were performed using NECI\cite{NECI_github}. DMRG benchmarks were generated with BLOCK\cite{Chan2002, Chan2004, Ghosh2008, Sharma2012, Olivares2015}.

We denote the SHCI threshold (controlling the size of $\mathcal{V}$) as $\epsilon$. For all SHCI calculations, the threshold for the perturbative stage (usually labelled $\epsilon_2$ or $\epsilon_{\textrm{PT}}$) is set to $10^{-10}$ Ha, which provides converged PT2 energies.

All calculations, both for SHCI and FCIQMC, use a many-particle basis of time-reversal symmetrized functions\cite{Smeyers1973}, rather than Slater determinants, although we use the term `determinant' for a many-particle basis state to be consistent with the common case. All of the theory presented is identical in either basis.

Note that we only consider estimators $\Evar$ (labelled `i-FCIQMC(SCI)') and $\EvarPT$ (labelled `i-FCIQMC(SCI)+PT2'). We do not consider projected energy estimators, such as $\Eref$. Although these have been more commonly used in FCIQMC, they are non-variational, making it challenging to fairly assess the accuracy of the sampled wave functions in different approximations. Moreover, in practice it is not uncommon to find that $\Evar$ is lower than $\Eref$.\cite{Overy2014, Blunt2017} Similarly, we also use the variational $\Evar$ estimator for all results from the original i-FCIQMC method. Although we do not consider projected energy estimators, we emphasize that these estimators also tend to be improved by the i-FCIQMC(SCI) approach.

\subsection{Details of molecules and orbitals used}
\label{sec:systems}

The systems studied and basis sets used are: formamide in a cc-pVDZ basis $(18\textrm{e}, 54\textrm{o})$; actone in a cc-pVDZ basis $(26\textrm{e}, 82\textrm{o})$; benzene in a cc-pVDZ basis $(30\textrm{e}, 108\textrm{o})$; butadiene in an ANO-L-VDZP$[3s2p1d]/[2s1p]$ basis, $(22\textrm{e},82\textrm{o})$ (as has been studied elsewhere\cite{Daday2012, Olivares2015, Chien2018, Guo2018_1, Blunt2018, Blunt2019}); hexacene in a cc-pVDZ basis, using the $(26\textrm{e}, 26\textrm{o})$ full $\pi$-valence space; 9,10-bis(phenylethynyl)anthracene (BPEA) in a cc-pVDZ basis, using the $(30\textrm{e}, 30\textrm{o})$ full $\pi$-valence space; and the water dimer in cc-pVDZ $(20\textrm{e}, 48\textrm{o})$, cc-pVTZ $(20\textrm{e}, 116\textrm{o})$ and cc-pVQZ $(20\textrm{e}, 230\textrm{o})$ basis sets.

For FCI problems we consider using both Hartree--Fock and optimized orbitals. Here, optimized orbitals were generated by the aHCISCF procedure, as described in Sec.~(\ref{sec:procedure_summary}). For the hexacene $(26\textrm{e}, 26\textrm{o})$ problem, CASSCF orbitals were generated both with and without active-active rotations optimized by HCISCF. For the BPEA $(30\textrm{e}, 30\textrm{o})$ problem, CASSCF orbitals with optimized active-active rotations were used for all calculations. The thresholds used by the SHCI solver in these optimizations are presented in supplementary material.

The geometries of formamide, acetone and benzene were taken from Ref.~(\onlinecite{Schreiber2008}). The geometry of butadiene was taken from Ref.~(\onlinecite{Daday2012}). The geometry of the water dimer was taken from Ref.~(\onlinecite{Lane2013}), optimized at the CCSDTQ/jun-cc-pVDZ level. The geometry of hexacene was taken from Ref.~(\onlinecite{Hachmann2007}), and the geometry of BPEA was optimized by PySCF (using the geomeTRIC library\cite{geometric}) at the B3LYP/6-31G\cite{Becke1993, Lee1988} level. For hexacene and BPEA, these geometries were then modified on the order of $\sim 10^{-4}$ \AA or less to exactly enforce $D_{2h}$ symmetry for subsequent calculations, which changed Hartree--Fock energies by less than $0.05$ $\mHa$. All geometries are presented in supplementary material.

\begin{figure}[t]
\includegraphics[width=\linewidth]{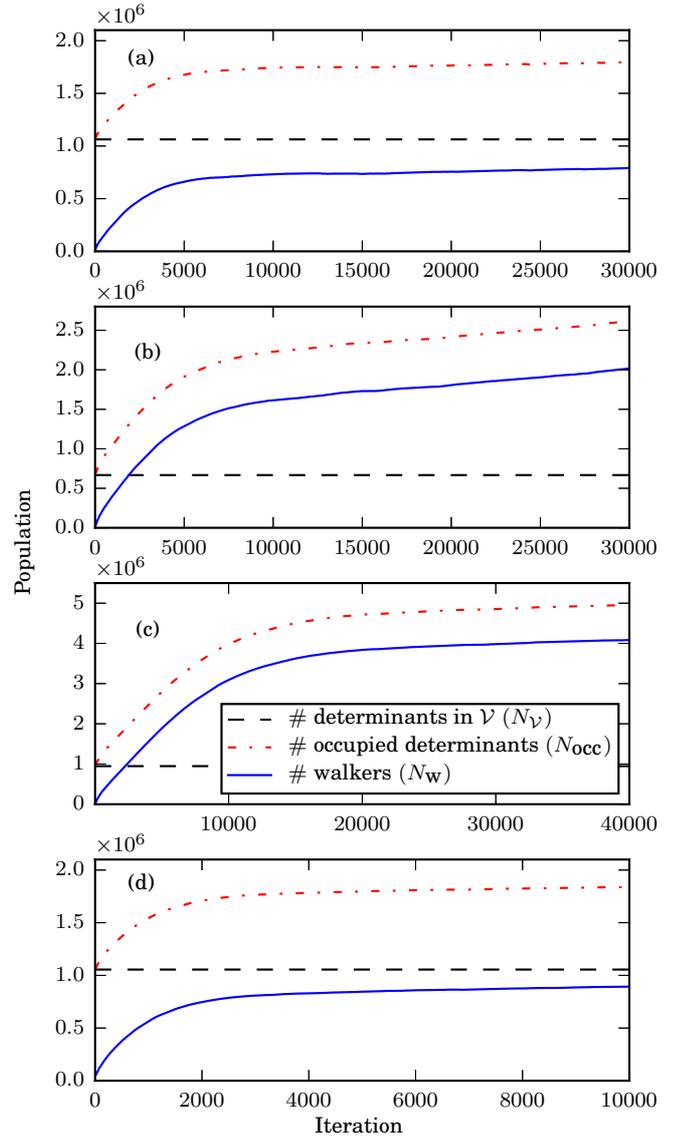}
\caption{Example population plateaus in i-FCIQMC(SCI) simulations. Results shown are the walker population (solid line), the size of SCI selected space, $\mathcal{V}$, (dashed), and the number of occupied determinants (dash-dot). The systems are (a) formamide $(18\textrm{e}, 54\textrm{o})$, (b) acetone $(26\textrm{e}, 82\textrm{o})$, (c) benzene $(30\textrm{e}, 108\textrm{o})$, and (d) hexacene $(26\textrm{e}, 26\textrm{o})$, using a cc-pVDZ basis in each case.}
\label{fig:plateaus}
\end{figure}

\begin{table*}
\begin{center}
{\footnotesize
\begin{tabular}{@{\extracolsep{4pt}}llccccccc@{}}
\hline
\hline
 & & & & \multicolumn{2}{c}{Plateau height (\# of walkers)} & \multicolumn{2}{c}{$\textrm{Plateau height}/N_{\mathcal{V}}$ } \\
\cline{5-6} \cline{7-8}
Basis     & Orbitals used  & $\epsilon$ (Ha) &   $N_{\mathcal{V}}$   & w/ semi-stoch.  & w/o semi-stoch.  & w/ semi-stoch.  & w/o semi-stoch. \\
\hline
cc-pVDZ   & HF                      & $1.5 \times 10^{-4}$ & $7.6 \times 10^{5}$ & $1.3 \times 10^{6}$ & $5.9 \times 10^{6}$ & $1.7$ & $7.7$ \\
cc-pVTZ   & HF                      & $2.0 \times 10^{-4}$ & $6.6 \times 10^{5}$ & $3.5 \times 10^{6}$ & $1.3 \times 10^{7}$ & $5.3$ & $19.7$ \\
cc-pVQZ   & HF                      & $4.0 \times 10^{-4}$ & $7.5 \times 10^{5}$ & $6.3 \times 10^{6}$ & $2.6 \times 10^{7}$ & $8.4$ & $34.7$ \\
\hline
cc-pVDZ   & Optimized               & $7.5 \times 10^{-5}$ & $7.5 \times 10^{5}$ & $1.2 \times 10^{5}$ & $1.1 \times 10^{6}$ & $0.16$ & $1.5$ \\
cc-pVTZ   & Optimized               & $7.4 \times 10^{-4}$ & $7.4 \times 10^{5}$ & $4.5 \times 10^{5}$ & $2.6 \times 10^{6}$ & $0.6$ & $3.5$ \\
\hline
\hline
\end{tabular}
}
\caption{Population plateau heights in i-FCIQMC(SCI) for the water dimer at equilibrium geometry. $N_{\mathcal{V}}$ denotes the number of determinants in $\mathcal{V}$. All electrons were correlated. We considered both Hartree--Fock orbitals and orbitals optimized by aHCISCF, both with and without the semi-stochastic adaptation. When the semi-stochastic adaptation was used, the deterministic space was set equal to $\mathcal{V}$. There is a modest increase in plateau height with basis set cardinal number. The semi-stochastic adaptation and optimized orbitals are both seen to reduce the sign problem.
}
\label{tab:plateau_heights}
\end{center}
\end{table*}

\begin{table}
\begin{center}
{\footnotesize
\begin{tabular}{@{\extracolsep{4pt}}lcc@{}}
\hline
\hline
Walker pop. ($\Nw$) &  $\Evar$  & $\%$ of energy from FOIS obtained \\
\hline
$10^3$  &  $-169.4362(6)$  &  $86.1\%$  \\
$10^4$  &  $-169.4402(2)$  &  $88.8\%$  \\
$10^5$  &  $-169.4467(1)$  &  $93.0\%$  \\
$10^6$  &  $-169.4530(2)$  &  $97.1\%$  \\
$10^7$  &  $-169.4567(3)$  &  $99.4\%$  \\
\hline
\hline
\end{tabular}
}
\caption{Results assessing the accuracy of the i-FCIQMC(SCI) approach as a function of walker population for a fixed selected/initiator space. The system is formamide in a cc-pVDZ basis. The SHCI selected space ($\mathcal{V}$) is of size $361$, obtained with $\epsilon = 0.02$ $\Ha$. The selected space together with its FOIS ($\mathcal{V}+\textrm{FOIS}$) is of size $\sim 3 \times 10^7$. The variational energy within $\mathcal{V}$ is $-169.3035$ $\Ha$. The variational energy within $\mathcal{V}+\textrm{FOIS}$ is $-169.4575$ $\Ha$. The percentage of the energy from the FOIS is defined as $\frac{(\Evar - E_{\mathcal{V}}) }{ (E_{\mathcal{V} + \textrm{FOIS}} - E_{\mathcal{V}}) } \times 100\%$.
}
\label{tab:interp}
\end{center}
\end{table}

\subsection{Plateau heights in the hybrid approach}
\label{sec:plateaus}

In i-FCIQMC, the number of initiators is always a small fraction of the total walker population, ensuring that no severe sign problem occurs. With a fixed initiator space, however, one would not expect to be able to use an arbitrarily small walker population. If the number of initiators is considerably larger than the walker population then we expect to observe a severe sign problem.

The severity of the sign problem in FCIQMC can typically be assessed by the height of the population plateau; only above this plateau can sensible results be obtained, so that the plateau essentially sets a minimum walker population for the simulation. It is therefore important to assess the typical plateau height in the i-FCIQMC(SCI) approach. As discussed in Section~\ref{sec:sci_iqmc}, the space variationally sampled in this approach (albeit with an approximate Hamiltonian) is $\mathcal{V} \oplus \mathcal{C}$. This is substantially larger than $\mathcal{V}$ in general, usually by several orders of magnitude. For the i-FCIQMC(SCI) approach to be worthwhile, the plateau height should be much lower than this space size.

Note that in the original FCIQMC method, plateaus represent the minimum walker population to stably sample the FCI solution\cite{Booth2009}; the FCI solution is obtained because no truncation is applied to the Hamiltonian. In i-FCIQMC(SCI), the population plateau also represents the minimum walker population to achieve a stable sampling, although the FCI solution is not obtained because the Hamiltonian is truncated. Nonetheless, the same theory\cite{Spencer2012} describes the plateau region and the algorithm behaves similarly in this region in both cases.

Examples of population plateaus are shown in Fig.~(\ref{fig:plateaus}). Each subplot shows the population dynamics of a simulation. The systems are formamide, acetone, benzene, and the $(26\textrm{e}, 26\textrm{o})$ full $\pi$-valence space of hexacene. The dashed line in each subplot shows the size of the selected/initiator space, $N_{\mathcal{V}}$, taken from SHCI. Both the walker population ($\Nw$) and also the number of occupied determinants ($\Nocc$) are shown. In the original algorithm, $\Nocc$ is always smaller than $\Nw$ because the minimum walker weight is $1$. With the semi-stochastic adaptation, however, determinants with amplitude less than $1$ are allowed in the deterministic space, so that $\Nocc > \Nw$ is typical. The initial value of $\Nocc$ is equal to $N_{\mathcal{V}}$, because we initialize from the SCI wave function. Note that the population begins to grow exponentially again after the plateau, although this behavior is not visible in Fig.~(\ref{fig:plateaus}) for the range of iterations shown.

It should be noted that the plateau height is in general slightly dependent on the simulation parameters. For example, it has been pointed out\cite{Spencer2012} that as the shift, $E_S$, approaches the exact ground-state energy, the plateau becomes lower but longer. We also observe the plateau height to depend slightly on the initial walker population. However, in general this variation is not significant. It should also be emphasized that one can bypass the plateau altogether by using an initial walker population above the plateau height. We only investigate plateaus here as a way of investigating the sign problem severity. Details of the SHCI and FCIQMC parameters used are given in the supplementary material.

For formamide and hexacene, the plateau height in $\Nw$ is actually \emph{below} the size of the initiator space, and the plateau in $\Nocc$ is within a factor of $2$ of this size. This plateau height then grows slowly with system size. For benzene, which has $\sim 3.8 \times 10^5$ basis states connected to the Hartree--Fock determinant, the plateau in $\Nocc$ is larger than the initiator space by only a factor of between $4$ and $5$ times, demonstrating the effectiveness of the sampling.

All of the systems considered in Fig.~(\ref{fig:plateaus}) use a cc-pVDZ basis set. However, the size of FOIS space relative to that of $\mathcal{V}$ grows quickly with the orbital basis set, so it is important to investigate larger basis sets. This is done in Table~(\ref{tab:plateau_heights}), where we have considered the water dimer in cc-pVDZ, cc-pVTZ and cc-pVQZ basis sets. Additionally, we have considered plateau heights both with and without the semi-stochastic adaptation, and with both Hartree--Fock and optimized orbitals. In each case, we chose $\epsilon$ in the SHCI calculation to give a selected space of size $\sim 7.5 \times 10^5$. Our first observation is that using optimized orbitals substantially reduces the sign problem, with the plateau height reduced by roughly an order of magnitude in each case. Similarly, using the semi-stochastic adaptation also reduces the plateau height. We used a deterministic space equal to $\mathcal{V}$, though using a smaller space is also possible, which will modify the plateau height accordingly. There is a general increase in plateau height with basis set size, as expected, although the increase is only a small factor for each increase in the basis set cardinal number. Thus, studying quadruple-$\zeta$ basis sets is reasonable with this approach. However, we were not able to perform orbital optimization for cc-pVQZ due to current memory limitations, which could pose an issue in using optimized orbitals for large systems.

It is therefore possible to sample a variational energy within $\mathcal{V} \oplus \mathcal{C}$, with a walker population that is a small factor of the size of $\mathcal{V}$, although this is certainly dependent on the system, basis, and simulation parameters. However, it should again be emphasized that the Hamiltonian in this approach is not $\bs{H}_B$, and therefore the accuracy needs considering. Results in Table~\ref{tab:interp} show the accuracy of this approach as a function of walker population with a fixed selected/initiator space. Here we consider the same formamide example as above. In order that we may solve for the exact lowest eigenstate of $\bs{H}_B$, we take a very small selected space of size $361$ from SHCI (with $\epsilon = 0.02$ $\Ha$). The dimension of $\mathcal{V} \oplus \mathcal{C}$ is then $\sim 3 \times 10^7$. The lowest eigenvalues of $\bs{H}_{\mathcal{V}}$ and $\bs{H}_B$ are $-169.3035$ $\Ha$ and $-169.4575$ $\Ha$, respectively. It can be seen that even with a walker population of only $10^3$, around $86\%$ of this energy difference is retrieved by the hybrid procedure, even using a variational energy estimator. As $\Nw$ is increased, energies tend to the lowest eigenvalue of $\bs{H}_B$, as justified in Section~\ref{sec:sci_iqmc}.

\subsection{Comparison with i-FCIQMC}
\label{sec:ifciqmc_comp}

\begin{figure}
\includegraphics{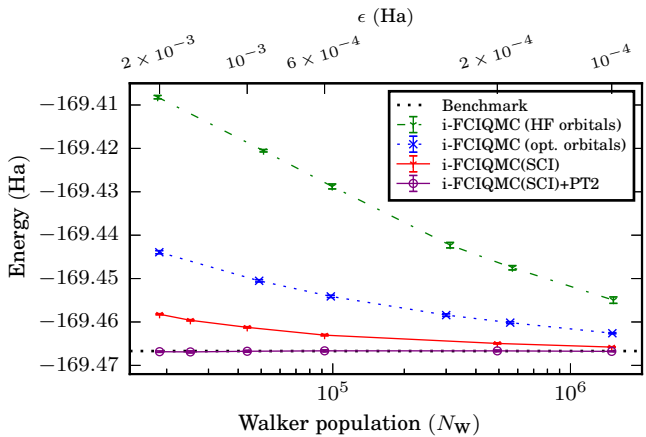}
\caption{Energies for formamide in a cc-pVDZ basis. For i-FCIQMC(SCI) results, walker populations equal to $\sim 1.5$ times the size of the selected spaces were used. The $\epsilon$ values for these SCI calculations are shown on the upper $X$-axis. i-FCIQMC simulations (where initiators are chosen by the population criterion) were then run with roughly the same walker populations for comparison. The benchmark is an extrapolated SCI+PT2 result which is essentially exact.}
\label{fig:formamide_ifciqmc}
\end{figure}

We now consider comparison with the previous i-FCIQMC approach, where initiators are defined by the population criterion. Fig.~(\ref{fig:formamide_ifciqmc}) shows results for formamide. We consider i-FCIQMC using both Hartree--Fock (HF) orbitals and optimized orbitals. These results are then compared to the hybrid approach where initiators are defined by the selected space from SHCI, with the corresponding SHCI parameters ($\epsilon$) shown on the upper $X$-axis. For this hybrid approach, walker populations of roughly $1.5$ times the size of $\mathcal{V}$ were used, which is sufficient to exceed the plateau. i-FCIQMC calculations were then run with the same walker populations for comparison.

\begin{figure}[t]
\includegraphics{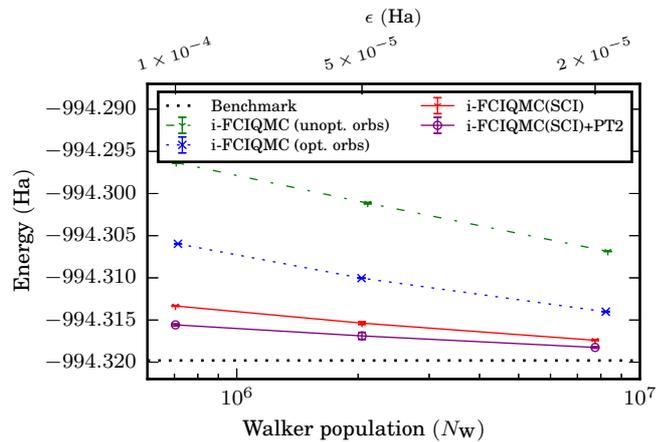}
\caption{Energies for hexacene in the $(26\textrm{e}, 26\textrm{o})$ full $\pi$-valence space. i-FCIQMC results (where initiators are chosen by the population criterion) were performed using CASSCF orbitals where active-active rotations had been optimized (labelled ``opt.'') and not optimized (``unopt.''). i-FCIQMC(SCI) results used the optimized orbitals only. The benchmark energy is $-994.31980$ $\Ha$ from DMRG with $M=4000$.}
\label{fig:hexacene_ifciqmc}
\end{figure}

From Fig.~(\ref{fig:formamide_ifciqmc}) it can be seen that optimized orbitals give substantially better energies than Hartree--Fock orbitals. This is expected, and has already been noted in SHCI\cite{Smith2017, Chien2018}, but it is nonetheless interesting to note the size of the improvement in FCIQMC. For example, with a small walker population of $\sim 1.8 \times 10^4$, the error in i-FCIQMC with HF orbitals is $\sim 59$ $\mHa$, which is reduced to $\sim 23$ $\mHa$ with optimized orbitals. Using the i-FCIQMC(SCI) approach (with an initiator space equal to the SHCI selected space with $\epsilon=0.002$ $\Ha$), this error is reduced to $8.5$ $\mHa$ for the same walker population. If the PT2 correction is included then this error is reduced to less than $0.2$ $\mHa$, with a statistical error of $0.17$ $\mHa$. i-FCIQMC(SCI) simulations used preconditioning\cite{Blunt2019} and a larger spawned vector list in order to reduce statistical errors on the PT2 correction. They also used a larger deterministic space for the semi-stochastic adaptation, requiring extra memory for the deterministic Hamiltonian. However, computational costs were otherwise very similar for all results of a given $\Nw$. Thus, even for this non-trivial system with $18$ electrons in $54$ orbitals, near-FCI accuracy can be obtained with a walker population of $\sim 1.8 \times 10^4$.

Hexacene is considered next, taking the $(26\textrm{e}, 26\textrm{o})$ full $\pi$-valence space. This is a more strongly correlated example, which has been studied by electronic structure methods several times previously\cite{Hachmann2007, Hajgato2009, Kurashige2014}. Results are shown in Fig.~(\ref{fig:hexacene_ifciqmc}). The benchmark energy is from DMRG with $M=4000$ renormalized states, which was verified against extrapolated SHCI. For i-FCIQMC results (using the population criterion) we use CASSCF orbitals both with active-active rotations performed (labelled ``opt.'') and not performed (``unopt.''), using the approach described in Sec.~(\ref{sec:systems}). The former optimized orbitals were then used for i-FCIQMC(SCI) results.

For i-FCIQMC(SCI), walker populations $\sim 2$ times greater than the size of the selected space were used. Simulations using the original initiator method were then performed at the same walker populations. Results show a similar trend to those for formamide. The variational energy is substantially reduced by using optimized orbitals, and further reduced using the hybrid approach to form the initiator space from SHCI. The PT2 correction is relatively small for this system, removing less than $50\%$ of error, presumably because the system is more strongly correlated.

\subsection{Comparison with SCI+PT2}
\label{sec:sci_comp}

\begin{figure}[t]
\includegraphics{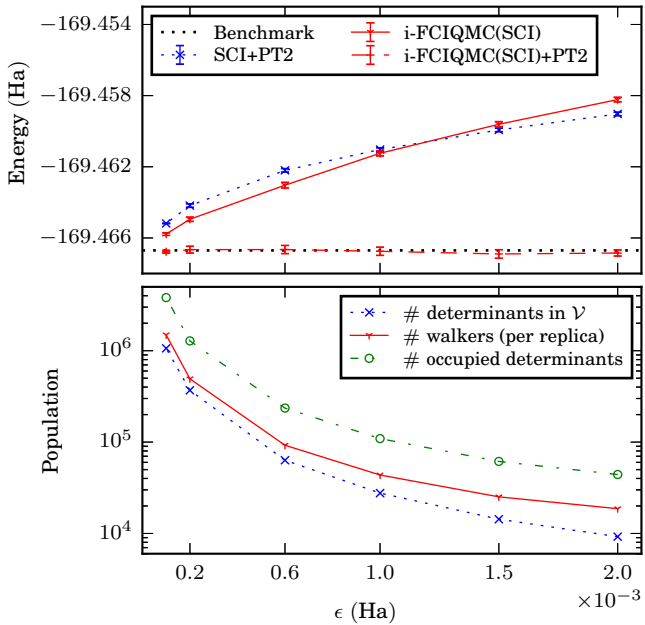}
\caption{Results for formamide in a cc-pVDZ basis. Results show SCI+PT2 using the SHCI approach, and i-FCIQMC(SCI) results using the same selected space, as labelled by the SHCI parameter, $\epsilon$. Top: Energies from SCI+PT2, i-FCIQMC(SCI) and i-FCIQMC(SCI)+PT2, as the SHCI parameter ($\epsilon$) is varied. Bottom: The size of the SHCI selected space compared to the walker population (per replica), and the total number of occupied determinants (across both replicas).}
\label{fig:formamide_sci_comp}
\end{figure}

We next consider comparison with SCI+PT2, in particular the semi-stochastic heat-bath CI (SHCI) method. The space sampled by the surviving spawns in i-FCIQMC(SCI) is identical to the FOIS sampled by the PT2 correction in SHCI. As such, it is not unreasonable to expect $\Evar$ in i-FCIQMC(SCI) to be comparable to the perturbatively-corrected energy in SHCI. However, the rejected spawns in the hybrid approach can also be used to form a perturbative correction in the second-order space, which one might expect to be substantially more accurate, particularly in weakly-correlated systems. In exchange for this improved accuracy, this QMC approach will be slower than SHCI for a common selected/initiator space.

We first consider the formamide molecule again, taking the same geometry, basis set and optimized orbitals as used in Sec.~(\ref{sec:ifciqmc_comp}). Results are presented in Fig.~(\ref{fig:formamide_sci_comp}). The $\epsilon$ values on the $X$-axis denote the thresholds used for SHCI, and the same selected spaces are used in the subsequent i-FCIQMC(SCI) calculations. Note that i-FCIQMC(SCI) and i-FCIQMC(SCI)+PT2 results are identical to those presented in Fig.~(\ref{fig:formamide_ifciqmc}). As expected, SCI+PT2 has similar accuracy to i-FCIQMC(SCI), although it should be noted that the latter estimate is rigorously variational (within statistical errors). Including the PT2 correction for i-FCIQMC(SCI) gives substantially more accurate results, as previously seen for this system in Fig.~(\ref{fig:formamide_ifciqmc}). Indeed, these results are all exact within statistical errors of $0.1 - 0.2$ $\mHa$, even with the largest $\epsilon$ value considered.

\begin{figure}[t]
\includegraphics{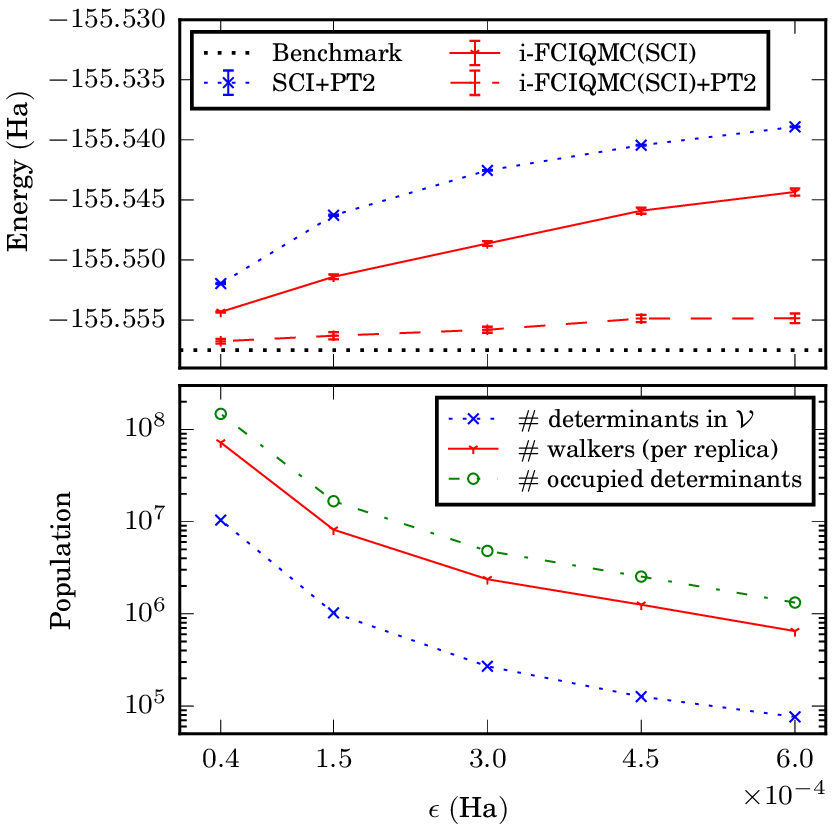}
\caption{Results for butadiene with the ANO-L-VDZP$[3s2p1d]/[2s1p]$ basis\cite{Daday2012,Olivares2015,Chien2018,Guo2018_1}. The active space is $(22\textrm{e},82\textrm{o})$. The benchmark is an extrapolated DMRG+PT2 result from Ref.~(\onlinecite{Guo2018_1}).}
\label{fig:butadiene_sci_comp}
\end{figure}

Next we consider the butadiene molecule. Results are presented in Fig.~(\ref{fig:butadiene_sci_comp}). For a given selected space, the hybrid approach is more accurate than SHCI here, although the walker population used to achieve this result is notably somewhat higher than the size of $\mathcal{V}$; walker populations used were as much as $10$ times larger than the size of the selected space used for this system. This larger walker population was chosen primarily to reduce statistical noise on the perturbative correction. As for formamide, including this PT2 correction to i-FCIQMC(SCI) results in much improved energies, removing $\sim 80\%$ of error for each $\epsilon$ value.

\begin{figure}[t]
\includegraphics{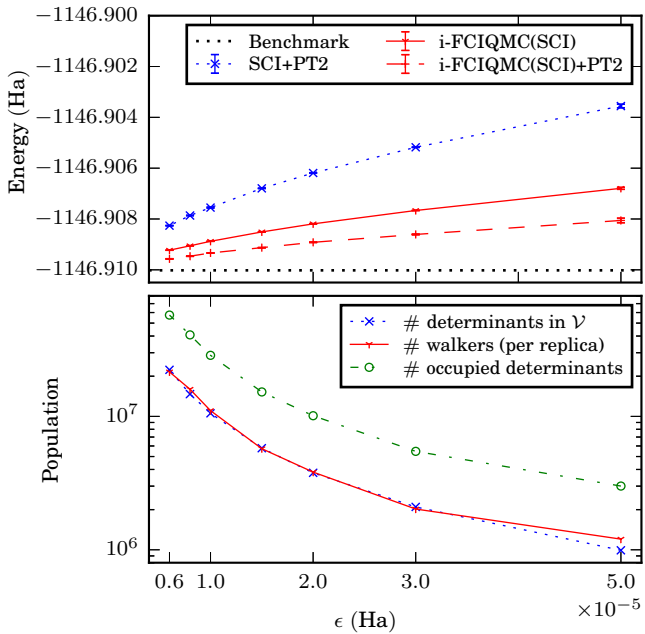}
\caption{Results for BPEA. The active space is the $(30\textrm{e},30\textrm{o})$ full $\pi$-valence space, and CASSCF orbitals were used, obtained by a prior aHCISCF calculation in a cc-pVDZ basis. The benchmark energy is $-1146.91002$ $\Ha$ from DMRG with $M=3000$.}
\label{fig:bpea_sci_comp}
\end{figure}

Lastly we consider BPEA. This molecule was recently experimentally suggested as a singlet fission chromophore\cite{Manna2018, Bae2018}. We do not assess the singlet fission status of BPEA here, saving this task for future work, but instead use BPEA as a further test system. We consider the full $\pi$-valence space, $(30\textrm{e},30\textrm{o})$. A CASSCF orbital optimization was performed including optimization of active-active rotations. The results in Fig.~(\ref{fig:bpea_sci_comp}) then demonstrate convergence to the exact CASCI solution with the resulting orbitals. The benchmark energy is from DMRG with $M=3000$, and is consistent with extrapolated SHCI results. i-FCIQMC(SCI) variational energies are seen to be more accurate than SCI+PT2 for a common selected space, here removing about $50\%$ of error. Similarly to results for hexacene, the PT2 correction is less successful than for weakly-correlated examples, but nonetheless a notable improvement. Furthermore, these perturbatively-corrected results can be used to successfully extrapolate to the exact limit, in an identical manner to the extrapolation of SCI+PT2 against variational SCI energies. For this system the walker populations used (per replica) were essentially identical to the corresponding selected space sizes, while the total number of occupied determinants (across both replicas) was larger by a factor of $\sim 3$ times.

We now consider the relative computational costs in SCI+PT2 and i-FCIQMC(SCI)+PT2 approaches. We specifically consider the SHCI approach to SCI+PT2 here. The lower subplots of Figs.~(\ref{fig:formamide_sci_comp}), (\ref{fig:butadiene_sci_comp}) and (\ref{fig:bpea_sci_comp}) can be used to assess the fundamental memory costs of each approach. In particular, SCI at least requires a small number of arrays the size of $\mathcal{V}$, whereas FCIQMC requires at least a small number of arrays the size of $\Nocc$. These results suggest that, for a given amount of memory, the hybrid approach can be used to obtain more accurate results, particularly for weakly correlated systems. It should be emphasized, however, that both methods typically store several other large arrays in order to improve their efficiency. In particular, SHCI typically stores the sparse Hamiltonian in $\mathcal{V}$, and uses the Davidson method, which requires holding several additional vectors. The PT2 correction can also be made substantially more efficient by using large memory arrays, first by performing a larger calculation in the deterministic PT2 step, and second by using larger batches in the stochastic PT2 step. Similarly, FCIQMC also typically stores other large arrays, including the sparse Hamiltonian for the semi-stochastic approach. Also, we recently showed that the statistical error on $\Delta E_2$ can be reduced more quickly\cite{Blunt2019} by performing $\Nspawn > 1$ spawns per walker. This requires setting the spawning array appropriately larger compared to the main walker array. For the above calculations for BPEA, formamide and butadiene, $\Nspawn$ was set between $10$ and $500$ (exact values are given in supplementary material). Therefore, the spawning arrays formed the main memory requirements in the above simulations. This approach speeds up the calculation of the PT2 correction by a factor of $\sim \sqrt{\Nspawn}$, as argued in Ref.~(\onlinecite{Blunt2019}), so that this not essential but certainly helpful. Timings are challenging to compare because they are highly implementation-dependent, and also highly-dependent on the simulation parameters used, including the amount of memory assigned for this task. However, SHCI is faster for a common selected space between the two approaches, usually by an order of magnitude or more on identical hardware. As such, the hybrid i-FCIQMC(SCI)+PT2 approach presented here can be seen as a complementary alternative to SCI+PT2 methods. The ability to exactly sample reduced density matrices\cite{Overy2014, Blunt2017} for the FOIS wave function can perhaps be seen as a particular benefit, which should make accurate estimation of properties more reliable. The hybrid approach also reduces the need for extrapolations. Moreover, FCIQMC-based approaches avoid the need to store an entire vector in $\mathcal{V}$ on each computing node, which ultimately limits the size of the selected space that can be used in SCI methods.

\section{Conclusion}
\label{sec:conclusion}

We have presented an approach to combining SCI and i-FCIQMC methods. By choosing the initiator space in i-FCIQMC to equal the selected space from a prior SCI calculation, the space sampled can be made substantially larger than in the current i-FCIQMC approach for a given walker population, thus improving the accuracy over the current method. This comes at the cost of reintroducing population plateaus into the method, although this perhaps suggests more efficient use of the sparse sampling in FCIQMC. Compared to i-FCIQMC, this approach gives systematically lower variational energies for a given walker population. It has also been shown that using orbitals optimized by aHCISCF greatly benefits this approach, both by improving accuracy and also by reducing the severity of the sign problem. We also find these orbitals to reduce the statistical error on $\Delta E_2$. Interestingly, using the semi-stochastic adaptation can reduce the severity of the sign problem further.

It has also been shown that the walker population needed to accurately sample the FOIS beyond $\mathcal{V}$ is, for all systems studied, only a small factor of the size of $\mathcal{V}$ itself, even when the FOIS is larger by several orders of magnitude. RDMs can then be constructed in this space, as already performed in FCIQMC\cite{Overy2014, Blunt2017}, which will be important for accurately calculating properties in future studies. The resulting variational energies from i-FCIQMC(SCI) wave functions are comparable, and sometimes lower than SCI+PT2 energies using the same selected space. Further including the PT2 correction to i-FCIQMC(SCI) can improve energies substantially, particularly for weakly correlated systems.

A current limiting factor to this approach is that the statistical error on the PT2 correction is often large, particularly due to the use of the replica trick. Currently, this requires large spawning arrays to improve the situation, and often extended running time. We are investigating alternative approaches to similar corrections that avoid using the replica trick, which we hope will improve this situation significantly. We also expect that the speed of FCIQMC can be significantly improved with excitation generators optimized for the preconditioned algorithm. With these developments, there is a hope of providing benchmark accuracy for a new range of systems in the near future.

\section{Supplementary Material}
Supplementary material includes details of the systems studies, including molecular geometries, orbital basis sets and the frozen core status of each molecule considered. Also presented are the SHCI thresholds used for aHCISCF orbital optimizations, and additional simulation parameters. Additional results for the formamide molecule are presented and discussed.

\begin{acknowledgments}
We are very grateful to St John's College, Cambridge for funding and supporting this work through a Research Fellowship. This study made use of the CSD3 Peta4-Skylake CPU cluster.
\end{acknowledgments}

%\bibliography{sci_iqmc}

\begin{thebibliography}{70}
\bibitem [{fn1()}]{fn1}%
  \BibitemOpen
  \href@noop {} {}\bibinfo {note} {In the NECI FCIQMC code used in this work, a
  determinant with amplitude $C_i$ will perform $\lfloor \abs{C_i} \rfloor$
  spawning attempts with walker weight $1$, and an extra attempt with
  probability $\abs{C_i} - \lfloor \abs{C_i} \rfloor$. This approach is
  identical within statistical errors.}\BibitemShut {Stop}%
\end{thebibliography}
%

\end{document}